\documentclass[conference]{IEEEtran}
\usepackage{blindtext}
\usepackage{graphicx}
\usepackage{multirow}
\usepackage{threeparttable}

\ifCLASSINFOpdf
\else
\fi
\usepackage[cmex10]{amsmath}
\usepackage{amsfonts}

\usepackage{url}
\begin{document}
%
% paper title
% Titles are generally capitalized except for words such as a, an, and, as,
% at, but, by, for, in, nor, of, on, or, the, to and up, which are usually
% not capitalized unless they are the first or last word of the title.
% Linebreaks \\ can be used within to get better formatting as desired.
% Do not put math or special symbols in the title.
\title{Optimal Dispatch of Electrified Autonomous Mobility on Demand Vehicles during Power Outages}

% % author names and affiliations
% % use a multiple column layout for up to three different
% % affiliations
% \author{\IEEEauthorblockN{Laurel Dunn and Sangjae Bae and Max Gardner and Colin Sheppard}
% \IEEEauthorblockA{Department of Civil \& Environmental Engineering\\
% University of California, Berkeley\\
% Berkeley, CA 30332--0250\\
% Email: lndunn@lbl.gov}
% \and
% \IEEEauthorblockN{Homer Simpson}
% \IEEEauthorblockA{Twentieth Century Fox\\
% Springfield, USA\\
% Email: homer@thesimpsons.com}
% \and
% \IEEEauthorblockN{James Kirk\\ and Montgomery Scott}
% \IEEEauthorblockA{Starfleet Academy\\
% San Francisco, California 96678--2391\\
% Telephone: (800) 555--1212\\
% Fax: (888) 555--1212}}

% conference papers do not typically use \thanks and this command
% is locked out in conference mode. If really needed, such as for
% the acknowledgment of grants, issue a \IEEEoverridecommandlockouts
% after \documentclass

% for over three affiliations, or if they all won't fit within the width
% of the page, use this alternative format:
% 
\author{
\IEEEauthorblockN{Colin Sheppard\IEEEauthorrefmark{1}\IEEEauthorrefmark{2}}
\IEEEauthorblockA{colin.sheppard@berkeley.edu}
\and
\IEEEauthorblockN{Laurel N. Dunn\IEEEauthorrefmark{1}\IEEEauthorrefmark{2}}
\IEEEauthorblockA{lndunn@lbl.gov}
\and
\IEEEauthorblockN{Sangjae Bae\IEEEauthorrefmark{1}\IEEEauthorrefmark{2}}
\IEEEauthorblockA{sbae@lbl.gov}
\and
\IEEEauthorblockN{Max Gardner\IEEEauthorrefmark{1}}
\IEEEauthorblockA{magardner@berkeley.edu}
\and
\IEEEauthorblockA{\IEEEauthorrefmark{1}Dept of Civil \& Environmental Engineering\\
University of California Berkeley,
Berkeley, CA 94720}
\and
\IEEEauthorblockA{\IEEEauthorrefmark{2}Lawrence Berkeley National Laboratory\\
Berkeley, CA 94720}}

% use for special paper notices
%\IEEEspecialpapernotice{(Invited Paper)}

% make the title area
\maketitle

% As a general rule, do not put math, special symbols or citations
% in the abstract
\begin{abstract}
The era of fully autonomous, electrified taxi fleets is rapidly approaching, and with it the opportunity to innovate myriad on-demand services that extend beyond the realm of human mobility. This project envisions a future where autonomous plug-in electric vehicle (PEV) fleets can be dispatched as both a taxi service and a source of on-demand power serving customers during power outages. We develop a PDE-based scheme to manage the optimal dispatch of an autonomous fleet to serve passengers and electric power demand during outages as an additional stream of revenue. We use real world power outage and taxi data from San Francisco for our case study, modeling the optimal dispatch of several fleet sizes over the course of one day; we examine both moderate and extreme outage scenarios. In the moderate scenario, the revenue earned serving power demand is negligible compared with revenue earned serving passenger trips. In the extreme scenario, supplying power accounts for between \$1 and \$2 million, amounting to between 32\% and 40\% more revenue than is earned serving mobility only, depending on fleet size. While the overall value of providing on-demand power depends on the frequency and severity of power outages, our results show that serving power demand during large-scale outages can provide a substantial value stream, comparable to the value to be earned providing grid services.
\end{abstract}

% no keywords

% For peer review papers, you can put extra information on the cover
% page as needed:
% \ifCLASSOPTIONpeerreview
% \begin{center} \bfseries EDICS Category: 3-BBND \end{center}
% \fi
%
% For peerreview papers, this IEEEtran command inserts a page break and
% creates the second title. It will be ignored for other modes.
\IEEEpeerreviewmaketitle

\section{Introduction}
% no \IEEEPARstart
\subsection{Motivation and Background}
Fully autonomous PEVs have tremendous potential to change the future of mobility. In particular, fleets of autonomous vehicles providing on-demand mobility services will likely play a major role in transportation systems. While the impact of these changes on travel demand is uncertain, it is clear that safety, energy efficiency, and cost of travel will be substantially improved in the future. It is also clear that autonomous on-demand fleets of PEVs will require continued innovation in methods for systems optimization and control.

Autonomous PEV fleets could play an important role in providing flexibility services to the future electric grid. Another opportunity for grid-connected PEVs to add value is to supply electricity to buildings experiencing power outages when utility customers are willing to pay more for energy to avoid incurring further damages (e.g., due to business closures or food waste). The current work examines the additional revenue available to a fleet of autonomous PEVs dispatched to provide both a mobility-on-demand service and backup power during outages.

\subsection{Relevant Literature}
The current personal vehicle ownership paradigm involves gross under-utilization of vehicles, as personal vehicles sit idle for most of the day. This under-utilization makes grid-connected PEV batteries an excellent source of load flexibility, as they can charge or discharge as needed while vehicles are not in use. Numerous studies examine the capabilities \cite{lefloch_pde_2016, lefloch_pde_2015, ota_autonomous_v2g, behrouz_2014} and economics \cite{ota_autonomous_v2g} of using electric vehicles to provide grid services. A more limited body of research examines opportunities for vehicles (specifically PHEVs) to provide backup power to homes during outages \cite{shin2016}.

Current trends suggest that the future of transportation is autonomous. Once autonomous vehicles are deployed at scale, the current paradigm of personal vehicle ownership is likely to change. A commercially operated fleet of autonomous PEVs will be more heavily utilized (and thus less flexible) than privately owned vehicles are today. However, centralized control can increase the magnitude and reliability of aggregate response when price signals are adequate. 

\subsection{Focus of this Study}
We propose a PDE-based approach, described in \cite{lefloch_pde_2016}, to simulate the optimal dispatch of autonomous on-demand PEVs serving time varying, spatially distributed demand for mobility (passenger trips) and backup power. The fleet is dispatched to maximize profit earned from serving both passenger trips and power. The revenue earned for each trip serviced or kWh provided depends on the origin and destination of the trip, and the location of the power outage. We consider several fleet sizes, examining differences in vehicle dispatch, revenue earned, and unserved demand for trips/power. Key contributions of this work include the geospatial modeling of vehicle mobility, charging \& discharging, and inclusion of backup power as an ancillary revenue stream.

\section{Technical Description}
\subsection{Modeling Aggregations of Autonomous Electric Vehicles}
\begin{table}[!htbp]
% increase table row spacing, adjust to taste
\renewcommand{\arraystretch}{1.3}
% if using array.sty, it might be a good idea to tweak the value of
% \extrarowheight as needed to properly center the text within the cells
\caption{Nomenclature}
\label{tab:symbols}
\centering
\def\colmargin{6.75cm}
% Some packages, such as MDW tools, offer better commands for making tables
% than the plain LaTeX2e tabular which is used here.
\begin{tabular}{ll}
\hline
\textbf{Symbol} & \textbf{Description}\\
\hline
$x$ & PEV Battery SOE ($dx=0.2$) \\
$t$ & Time ($dt=10 min$) \\
$N_n$ & Number of nodes ($3$)\\
$N_b$ & Number of spatial bins\\
$E_{max}$ & Battery energy capacity ($10 kWh$) \\
$\eta$ & Power conversion efficiency during charging ($0.86$) \cite{veic_2013} \\
$u_i(x,t)$ & Density of charging PEVs in node $i$\\
$v_i(x,t)$ & Density of idle PEVs in node $i$\\
$w_i(x,t)$ & Density of discharging PEVs in node $i$\\
$\sigma_{I_i \rightarrow C_i}(x,t)$ & \parbox[t]{\colmargin}{ \raggedright Flow of PEVs in node $i$ from Idle to Charging} \\
$\sigma_{I_i \rightarrow D_i}(x,t)$ & \parbox[t]{\colmargin}{Flow of PEVs in node $i$ from Idle to Discharging} \\
$\sigma_{I_i \rightarrow I_j}^o(x,t)$ & \parbox[t]{\colmargin}{Flow of PEVs from Idle state of node $i$ to Idle state of node $j$ without passengers} \\
$\sigma_{I_i \rightarrow I_j}^\prime(x,t)$ & \parbox[t]{\colmargin}{ Flow of PEVs from Idle state of node $i$ to Idle state of node $j$ with passengers} \\
$q_C(x,t)$ & Instantaneous charging power \\
$q_D(x,t)$ & Instantaneous discharging power \\
$Z$ & Set of Transportation Network Nodes (I, II, IV) (see Figure \ref{fig:sf}) \\
$T$ & Time horizon of the optimization ($50 min$) \\
$\rho_{dis}(i)$ & Price of servicing load during power outages by node(\$/kWh) \\
$\rho_{mob}(i,j)$ & \parbox[t]{\colmargin}{ Price of servicing mobility demand from node $i$ to node $j$ (\$/trip/minute)} \\
\hline
\end{tabular}
\end{table}
We adopt and extend the scheme developed by \cite{lefloch_pde_2016} for tracking and controlling an aggregation of electric vehicles. The core advantage of the scheme is the recognition that in an autonomous PEV fleet, only the location of vehicles and their state of charge are critical to know at any point in time. Instead of representing individual vehicles explicitly and developing a combinatorial approach to control, we aggregate all vehicles in a node and represent the aggregate distribution of vehicle state of energy (SOE). Vehicles in any node $i$ can be in one of three states: charging, idle, or discharging, which we represent by the state variables $u_i(x,t)$, $v_i(x,t)$, and $w_i(x,t)$, respectively. The system is then characterized by the following coupled partial differential equations (see Table \ref{tab:symbols} for further nomenclature):
\begin{eqnarray*}
    \frac{\partial u_i}{\partial t}(x,t) &=& -\frac{\partial}{\partial x}\left[ q_C(x) u_i(x,t) \right] + \sigma_{I_i \rightarrow C_i}(x,t) \\
    \frac{\partial v_i}{\partial t}(x,t) &=& \sum_{j\in Z} \left[ \sigma_{I_i \leftarrow I_j}^\prime(x,t) + \sigma_{I_i \leftarrow I_j}^o(x,t) \right. \\
    & & ~~~ \left. - \sigma_{I_i \rightarrow I_j}^\prime(x,t) - \sigma_{I_i \rightarrow I_j}^o(x,t) \right] \\
    && ~~~ - \sigma_{I_i \rightarrow C_i}(x,t) - \sigma_{I_i \rightarrow D_i}(x,t) \\
    \frac{\partial w_i}{\partial t}(x,t) &=& -\frac{\partial}{\partial x}\left[ q_D(x) w_i(x,t) \right] + \sigma_{I_i \rightarrow D_i}(x,t) 
\end{eqnarray*}
Where: 
\begin{eqnarray*}
q_C(x) &=& \frac{7}{E_{max}}\eta\frac{1}{60} \\
q_D(x) &=& \frac{-7}{E_{max}}\frac{1}{60}
\end{eqnarray*}
The equations make use of an advection term (when the time derivative is linearly related to the spatial derivative) to represent how SOE changes over time for vehicles in the charging or discharging states, with SOE advecting toward 1 or 0, respectively. The model is spatially disaggregated, so the three PDEs are repeated for every node in the system and indexed by $i$.

Flow terms $\sigma_{I_i \rightarrow C_i}(x,t)$ and $\sigma_{I_i \rightarrow D_i}(x,t)$ capture the transport of vehicles between the SOE curves for each state at a particular node. Additional flow terms capture transport between the Idle curves of different nodes. For a given node $i$ and any other node $j$, four separate terms represent trips with/without passengers ($\sigma^\prime$ and $\sigma^o$ respectively) and departing from/arriving at the node ($\sigma_{I_i \rightarrow I_j}$ and $\sigma_{I_j \leftarrow I_i}$ respectively). Collectively, all terms denoted by $\sigma$ are decision variables, where departures and arrivals are coupled via the constraints (as discussed below).

The inter-nodal flow terms are constrained such that departures from a node $i$ to node $j$ are equivalent to the arrivals of vehicles from $i$ to $j$ at a future time and with a lower SOE, corresponding to the travel time and energy requirements of that trip, as specified in Table \ref{tab:flow_constraints}. The distinction between trips with and without passengers becomes critical in the context of the economic optimization that places monetary value on transporting people but not on moving empty vehicles. Though we do account for the energy costs associated with moving empty vehicles, we do not consider the costs of any congestion these vehicles may cause.

\subsection{Optimization Formulation}
\subsubsection{Objective}
The objective of the optimization is to maximize the operational profit of dispatching the fleet of autonomous on-demand PEVs:
\begin{eqnarray*}
  \max_{\substack{\sigma_{I_i \rightarrow C_i} \\
    \sigma_{I_i \rightarrow D_i} \\ 
    \sigma_{I_i \rightarrow I_j}}}
    K &=& \sum_{i\in Z} \int_{t=0}^{T} \left[ \frac{\rho_{dis}(i)}{60} Q_{dis,i}(t) + \right. \\ 
      && \left. \sum_{j\in Z}\rho_{mob}(i,j)Q_{mob,i,j}(t)  - \frac{C}{60}Q_{ch,i}(t) \right]dt\\
    Q_{dis,i}(t) & = & \int_{0}^{1} 7 w_i(x,t) dx \\
    Q_{ch,i}(t) & = & \int_{0}^{1} 7 u_i(x,t) dx \\
    Q_{mob,i,j}(t) & = & \int_{0}^{1}\left( \sigma_{I_i \rightarrow I_j}^\prime(x,t) \right)dx
\end{eqnarray*}
Where $\rho_{mob}(i,j)$, $\rho_{dis}(i)$, and $C$ are the fares charged to passengers, the price charged to serve load during outages, and the cost to purchase electricity from the grid, respectively. The constant 60 converts kWh to kW-minutes, and the constant 7 is the charging and discharging rate of each vehicle consistent with current charging/discharging rates of Level 2 chargers.

\subsubsection{Constraints}
The equations of state are discretized using a first-order upwind scheme for numerically solving hyperbolic PDEs. They appear in the formulation as a set of equality constraints. Additional constraints limit the flow of vehicles between nodes to within realistic bounds, and ensure the overall conservation of vehicles in the system.

Firstly, we constrain the size of the flows between states $u$, $v$, and $w$ to be no greater than the number of vehicles in those states:
\begin{eqnarray*}
    -\sigma_{I_i \rightarrow C_i}(x,t) & \le & u_i(x,t) / \Delta t \\
  \left\{ \sigma_{I_i \rightarrow C_i}(x,t) + \sigma_{I_i \rightarrow D_i}(x,t) \right. && \\
  + \sigma_{I_i \rightarrow I_j}^\prime(x,t) + \sigma_{I_i \rightarrow I_j}^o(x,t) && \\
  \left. - \sigma_{I_i \leftarrow I_j}^\prime(x,t) - \sigma_{I_i \leftarrow I_j}^o(x,t) \right\} & \le & v_i(x,t) / \Delta t \\
    -\sigma_{I_i \rightarrow D_i}(x,t) & \le & w_i(x,t) / \Delta t
\end{eqnarray*}
We also require that as charging vehicles reach an SOE of 1 or as discharging vehicles reach an SOE of 0, they immediately flow to the Idle state.
\begin{eqnarray*}
    -\sigma_{I_i \rightarrow C_i}(1,t) & = & u_i(1,t) / \Delta t \\
    -\sigma_{I_i \rightarrow D_i}(0,t) & = & w_i(0,t) / \Delta t
\end{eqnarray*}
Next, we require that trips be conserved between origin-destination pairs, where arrivals are shifted to a later time step and a lower SOE, based on the time ($\Delta t$) and energy ($\Delta x$) requirements of the trip.
\begin{eqnarray*}
    \sigma_{I_i \rightarrow I_j}^\prime(x,t) & = & \sigma_{I_j \leftarrow I_i}^\prime(x - \Delta x_{i,j},t + \Delta t_{i,j}) \\
     \sigma_{I_i \rightarrow I_j}^o(x,t) & = & \sigma_{I_j \leftarrow I_i}^o(x - \Delta x_{i,j},t + \Delta t_{i,j})
\end{eqnarray*}
\begin{equation*}
\{(i,j) \in Z \times Z\} 
\end{equation*}
The values of $\Delta x$ and $\Delta t$ for each node (I, II and IV shown in Figure \ref{fig:sf}) are derived from historic taxi mobility and fare datasets such as \cite{mobility2009} and \cite{fares2017}, respectively. We assume a decline in personal vehicle ownership accompanies deployment of autonomous vehicles. We account for increasing reliance on mobility-on-demand services by scaling travel demand by a factor of 10 relative to 2012. We took the average trip durations and trip distances for trips from each node $i$ to each node $j$, scaling the average distance by 5.05 km/kWh to derive $\Delta x_{i,j}$ and taking the average time as $\Delta t_{i,j}$. The derived values are shown in Table \ref{tab:flow_constraints}.
\begin{table}[!htbp]
    % increase table row spacing, adjust to taste
    \renewcommand{\arraystretch}{1}
    % if using array.sty, it might be a good idea to tweak the value of
    % \extrarowheight as needed to properly center the text within the cells
    \caption{Flow Constraints}
    \label{tab:flow_constraints}
    \centering
    \def\colmargin{6.75cm}
    % Some packages, such as MDW tools, offer better commands for making tables
    % than the plain LaTeX2e tabular which is used here.
    \begin{tabular}{lll}
    \hline
    \textbf{Node Flows ($i$ $\rightarrow$ $j$)} & \textbf{Derived} $\Delta x$ (kWh) & \textbf{Derived} $\Delta t$ (s) \\
    \hline
    I$\rightarrow$I & 0.42  & 476  \\
    I$\rightarrow$II & 0.82  & 792  \\
    I$\rightarrow$IV & 0.93  & 1000  \\
    II$\rightarrow$I & 0.84  & 760  \\
    II$\rightarrow$II & 0.38  & 489  \\
    II$\rightarrow$IV & 0.77  & 698  \\
    IV$\rightarrow$I & 0.93  & 956  \\
    IV$\rightarrow$II & 0.77  & 725  \\
    IV$\rightarrow$IV & 0.37  & 403  \\
    \hline
    \end{tabular}
\end{table}
Vehicle dispatch is constrained such that the number of vehicles servicing passenger trips or power demand cannot exceed mobility and power demand at that time step.
\begin{eqnarray*}
   Q_{dis,i}(t) & \le & D_{dis,i}(t) \\
   Q_{mob,i,j}(t) & \le & D_{mob,i,j}(t) \\
\end{eqnarray*}
The demands $D_{dis,i}$ and $D_{mob,i,j}$ are exogenously defined; derivation of $D_{dis,i}$ is described below. The choice of inequality constraints when constraining $Q_{dis,i}$ and $Q_{mob,i,j}$ serves three purposes: 1) it allows the solution of the optimization to prioritize between serving the two types of demand; 2) it enables simulations where the fleet of vehicles is not sized to meet the peak demand in the system; and 3) it allows the system to be used in an application where power outages occur spontanteously and without foresight.

Finally, we require that the vehicle batteries have sufficient energy to make trips:
\begin{eqnarray*}
    \sigma_{I_i \rightarrow I_j}^\prime(x,t) & = & 0, ~~~ x < \Delta x_{i,j} \\
    \sigma_{I_i \rightarrow I_j}^o(x,t) & = & 0, ~~~ x < \Delta x_{i,j}
\end{eqnarray*}

Both the objective function and constraints are linear, making this a linear program. We have implemented the problem in R and use lp\_solve (an implementation of the simplex method) to find the optimal solution at each time step.

\subsection{Application}
\subsubsection{Spatial Discretization}
\begin{figure}[!htbp]
  \begin{center}
  \includegraphics[width=0.6\linewidth]{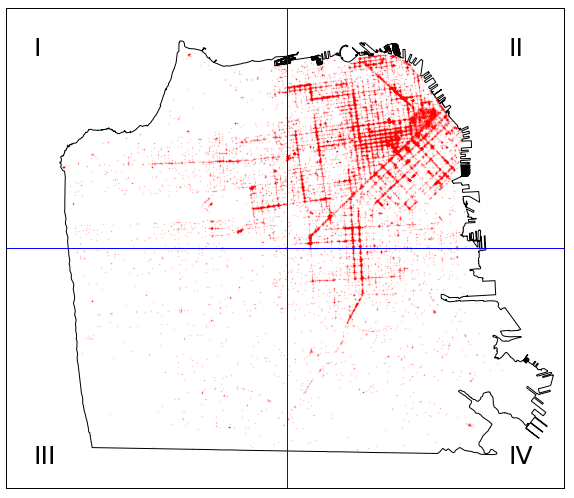}
  \end{center}
  \vspace{-5pt}
  \caption{We divide San Francisco into 4 equal-area nodes. Origin and destination of taxi trips are shown as red dots. Node III is excluded from the simulation due to limited demand.}
  \label{fig:sf}
\end{figure}

We divided the City of San Francisco, CA into a simplified 4-zone, equal-area network illustrated in Figure \ref{fig:sf}. As described above, we construct a mobility scenario to characterize demand for passenger trips to/from/within each node, travel time beteen nodes, and taxi fares. Below we describe how power outages are characterized from real world data. We observe very little demand for mobility and few outages in Node III. Thus due to additional computational complexity of modeling a four node system, we exclude Node III from our analysis.
\subsubsection{Demand for Backup Power}
We estimate the magnitude and location of power outages using historic outage data collected from the Pacific Gas \& Electric Company website. The data collected include the number and spatial distribution of power outages in the region over time; we aggregate outages spatially by node. We estimate the magnitude of unserved load based on the number of customers affected, the expected number of customers by customer type (i.e., residential, commercial, industrial), and average power demand by customer type (as reported in EIA form 861). In the current implementation, we assume demand to be the same accross all customers of a particular type and constant throughout the day. We use local population and economic census data to estimate the distribution of customers by type in each node.

\begin{figure}[!htbp]
  \begin{center}
  \includegraphics[width=0.8\linewidth]{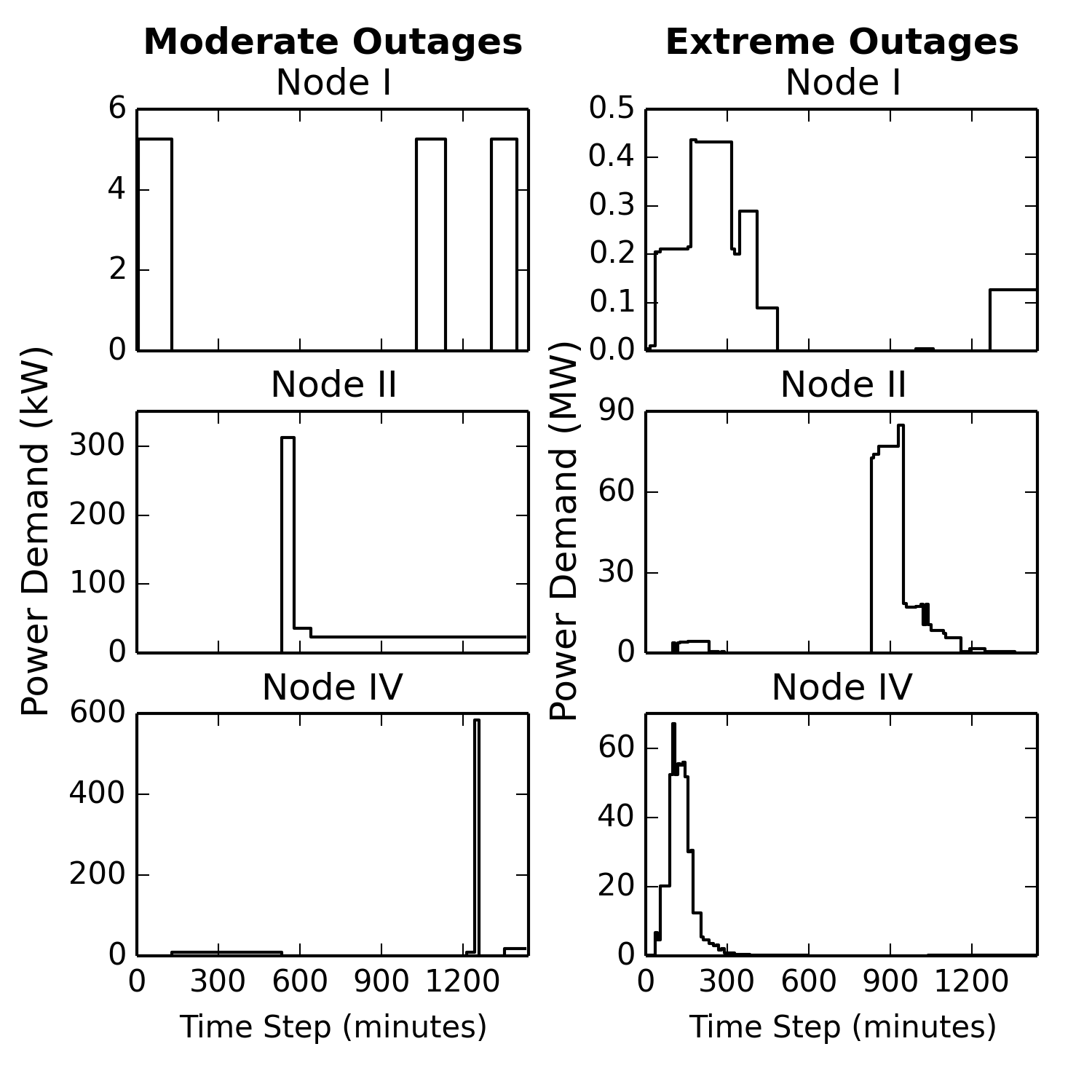}
  \end{center}
  \vspace{-5pt}
  \caption{Power demand at each node (I, II, IV) in the Moderate (left) and Extreme (right) outage scenarios, reprsented by September 29, 2014 and December 31, 2014, respectively. For readability, demand is reported in kWh in the Moderate scenario, and in MWh in the Extreme scenario.}
  \label{fig:power_demand}
\end{figure}

We examine two days of outage data, including one extreme outage scenario (December 31, 2014) and one moderate outage scenario (September 29, 2014). Figure \ref{fig:power_demand} shows the estimated power demand at each node for both scenarios. We highlight that demand in the Extreme outage scenario exceeds demand in the Moderate outage scenario by two orders of magnitude.
Finally, we estimate the value of providing on-demand backup power by computing the cost of damages incurred due to outages in each node for both outage scenarios. We use the Interruption Cost Estimate (ICE) Calculator to do so \cite{ice_calculator_2015}. Table \ref{tab:outage_costs} gives the estimated value of backup power in each node for the two outage scenarios in \$/kWh and \$/$\Delta t$ (where $\Delta t$ is 10 minutes). Although power demand is much higher in the Extreme outages scenario, the cost per kWh is greater in the Moderate outages scenario.
\begin{table}[!htbp]
    % increase table row spacing, adjust to taste
    \renewcommand{\arraystretch}{1}
    % if using array.sty, it might be a good idea to tweak the value of
    % \extrarowheight as needed to properly center the text within the cells
    \caption{Cost of power outages in each node for Extreme and Moderate outage scenarios per kWh delivered, and per time step (10 minutes).}
    \label{tab:outage_costs}
    \centering
    \def\colmargin{6.75cm}
    % Some packages, such as MDW tools, offer better commands for making tables
    % than the plain LaTeX2e tabular which is used here.
    \begin{tabular}{lllll}
    \hline
    \multirow{2}{*}{\textbf{Node ($i$)}} & \multicolumn{2}{c}{\textbf{Extreme}} & \multicolumn{2}{c}{\textbf{Moderate}} \\
     & (\$/kWh) & (\$/$\Delta t$) & (\$/kWh) & (\$/$\Delta t$) \\
    \hline
    I  & 20 & 23 & 14 & 16 \\
    II & 9  & 11 & 32 & 37 \\
    IV & 15 & 18 & 46 & 54 \\
    \hline
    \end{tabular}
\end{table}
For comparison, Table \ref{tab:trip_costs} lists the fares associated with passenger trips to and from each node in terms of dollars per unit energy consumed (\$/kWh) and dollars per unit time (\$/$\Delta t$). We highlight that the value earned per kWh serving passenger trips is remarkably similar to the value earned per kWh of power demand served.
\begin{table}[!htbp]
    % increase table row spacing, adjust to taste
    \renewcommand{\arraystretch}{1}
    % if using array.sty, it might be a good idea to tweak the value of
    % \extrarowheight as needed to properly center the text within the cells
    \caption{Cost of passenger trips per unit energy and per unit time for each origin-destination pair.}
    \label{tab:trip_costs}
    \centering
    \def\colmargin{6.75cm}
    % Some packages, such as MDW tools, offer better commands for making tables
    % than the plain LaTeX2e tabular which is used here.
    \begin{tabular}{llll}
    \hline
    \multirow{2}{*}{\textbf{Origin}} & \multirow{2}{*}{\textbf{Destination}} & \multicolumn{2}{c}{\textbf{Cost}} \\
     &  & \$/kWh & \$/$\Delta t$ \\
    \hline
    I  & I  & 25 & 11 \\
    I  & II & 19 & 8 \\
    I  & IV & 20 & 9 \\
    II & I  & 18 & 8 \\
    II & II & 26 & 10 \\
    II & IV & 19 & 7 \\
    IV & I  & 20 & 9 \\
    IV & II & 19 & 7 \\
    IV & IV & 24 & 9 \\
    \hline
    \end{tabular}
\end{table}
\section{Results}
We present simulation results for the two outage scenarios with various fleet sizes, including 7,500, 10,000 and 15,000 vehicles for the Moderate outage scenario, and 7,500, 15,000 and 40,000 vehicles for the Extreme outage scenario. The following sections detail the results. We highlight the revenue earned in different scenarios, and differences in dispatch among different fleet sizes.

\subsection{Revenue}
\begin{figure}[!htbp]
  \begin{center}
  \includegraphics[width=0.8\linewidth]{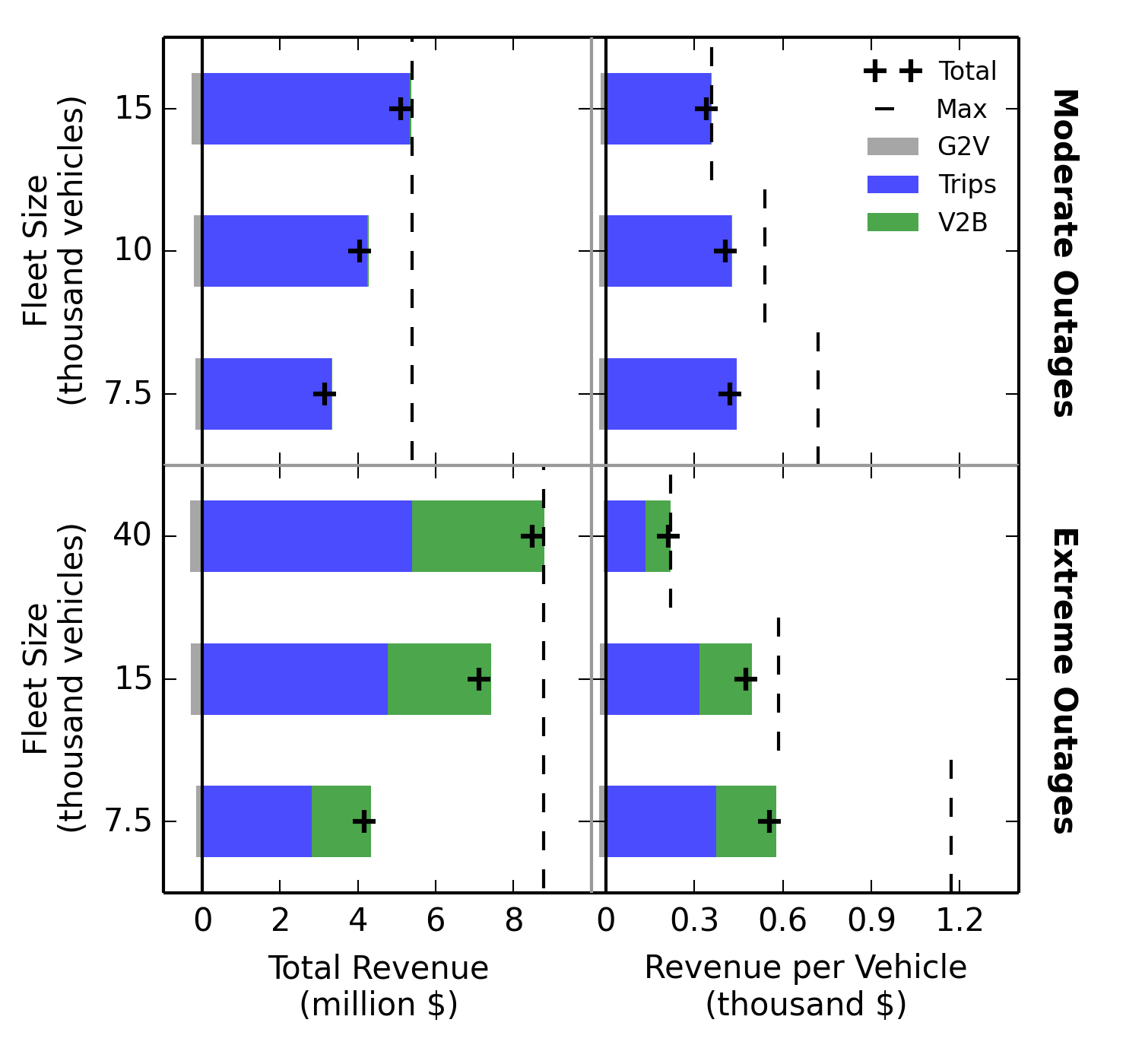}
  \end{center}
  \vspace{-5pt}
  \caption{Revenue earned by entire fleet (left) and per vehicle (right) in the Moderate (top) and Extreme (bottom) outage scenarios. Revenue components include: cost to charge (G2V), revenue earned serving passenger trips (Trips), and revenue earned serving power demand (V2B). The total revenue (Total) and maximum possible revenue (Max) are also shown.}
  \label{fig:revenue_bar}
\end{figure}

Figure \ref{fig:revenue_bar} presents the revenue earned in each scenario by the entire fleet and per vehicle. Contributors to overall revenue include: the cost to charge (G2V), revenue earned serving trips (Trips), and revenue earned serving power demand (V2B). The total revenue earned (Total) in each scenario and maximum possible revenue (Max) are also shown. The maximum possible revenue includes servicing all passenger trips and all power demand, with no charing costs.
Charging costs are almost negligible compared with the revenue earned because the cost of charging (0.25 \$/kWh) is small compared with the revenue earned serving power and mobility demand (see Tables \ref{tab:outage_costs} and \ref{tab:trip_costs}).

\subsection{Fleet Size and Vehicle Dispatch}
\begin{figure}[!htbp]
  \begin{center}
  \includegraphics[width=\linewidth]{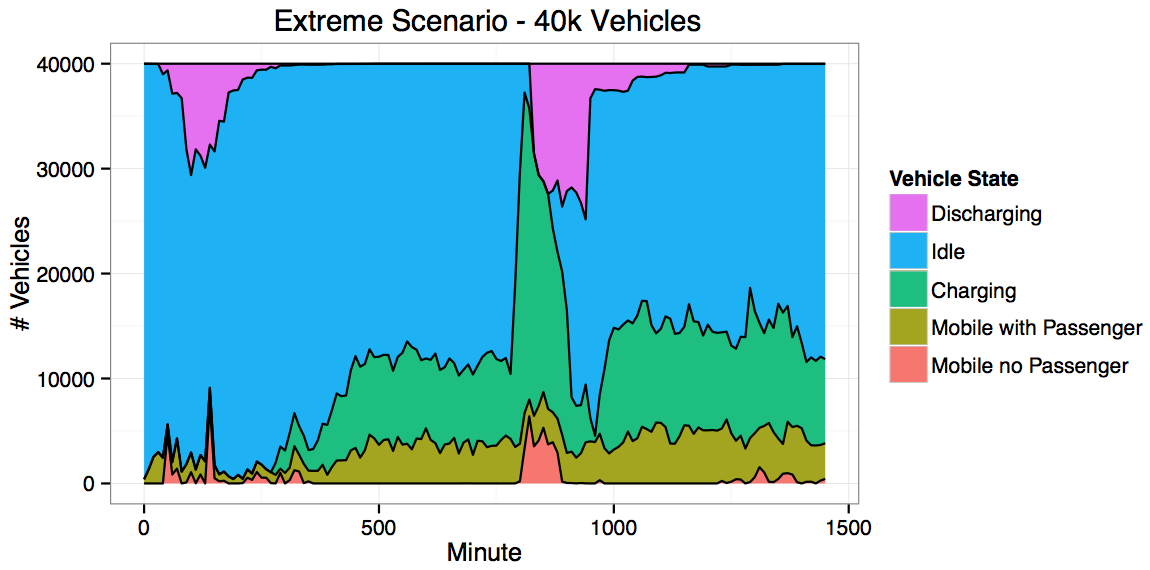}
  \end{center}
  \vspace{-5pt}
  \caption{Number of vehicles in each state at each time step in the Extreme outages scenario with a 40,000 vehicle fleet. States include: in transit with and without passengers, charging, discharging, and idle.}
  \label{fig:time_extreme_40k}
\end{figure}

\begin{figure}[!htbp]
  \begin{center}
  \includegraphics[width=\linewidth]{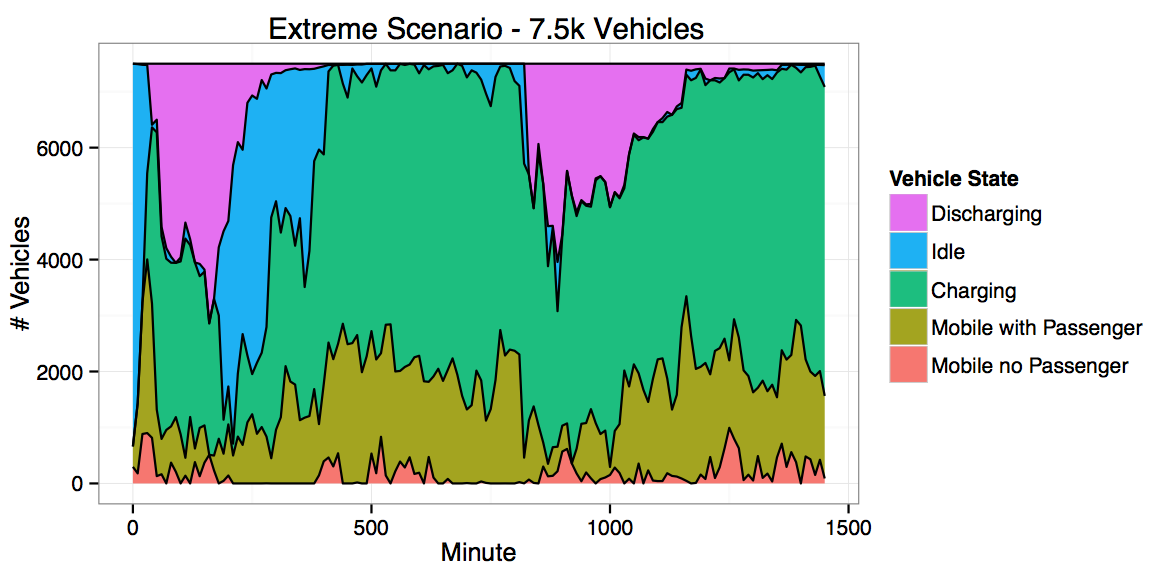}
  \end{center}
  \vspace{-5pt}
  \caption{Number of vehicles in each state at each time step in the Extreme outages scenario with a 7,500 vehicle fleet. States include: in transit with and without passengers, charging, discharging, and idle.}
  \label{fig:time_extreme_7500}
\end{figure}

Next we consider the benefits and drawbacks of different fleet sizes. Nearly all demand for mobility and power can be served with a 40,000 vehicle fleet in the Extreme scenario, and a 15,000 vehicle fleet in the Moderate scneario. Figures \ref{fig:time_extreme_40k} and \ref{fig:time_extreme_7500} show the number of vehicles in each state in the Extreme outages scenario with 40,000 and 7,500 vehicles. States include: discharging, idle, charging, and in transit both with and without passengers. 
Figure \ref{fig:time_extreme_40k} reveals that a 40,000 vehicle fleet spends most of the simulation in the idle state; the fleet is only fully utilized between 800 and 900 seconds when power demand peaks. Low revenue per vehicle in Figure \ref{fig:revenue_bar} provides further evidence that the 40,000 vehicle fleet is under-utilized. On the other hand, the 7,500 vehicle fleet in Figure \ref{fig:time_extreme_7500} earns less revenue overall, but spends very little time in the idle state. In fact, the vehicles spend more time charging than in any other state; faster charging infrastructure would increase fleet utilization. Future work will evaluate fast-charging infrastructure as an alternative to increasing the fleet size.
\section{Discussion}
To determine whether on-demand backup power provides a substantial value stream for the fleet, we consider the relative frequency of Extreme and Moderate outage days and the marginal revenue earned by serving power demand in addition to passenger trips. To do so, we compute the marginal annual revenue earned serving both power and mobility demand compared with serving mobility only. We examine several scenarios for the number of Extreme verses Moderate outage days in a year. We treat the Moderate outages scenario as a mobility-only scenario, as the revenue earned serving power demand in that scenario is negligible. The results are summarized in Table \ref{tab:annual_scenarios}.

\begin{threeparttable}
    % increase table row spacing, adjust to taste
    \renewcommand{\arraystretch}{1}
    % if using array.sty, it might be a good idea to tweak the value of
    % \extrarowheight as needed to properly center the text within the cells
    \caption{Increase in annual revenue from serving power demand in addition to mobility for 7,500 and 15,000 vehicle fleets with a range of scenarios regarding the number of Extreme outage days in the year.}
    \label{tab:annual_scenarios}
    \centering
    \def\colmargin{6.75cm}
    % Some packages, such as MDW tools, offer better commands for making tables
    % than the plain LaTeX2e tabular which is used here.
    \footnotesize{
    \begin{tabular}{p{1cm}p{1cm}p{1cm}p{1cm}p{1cm}}
    \hline
    \multirow{2}{1cm}{\textbf{Extreme Days}} & \multicolumn{2}{c}{\textbf{New Revenue} (\$/year/vehicle)} & \multicolumn{2}{c}{\textbf{Percent Increase} (\%)} \\
    &7,500 & 15,000 & 7,500 & 15,000 \\
    \hline
    10 & 1400 & 2000 & 0.9 & 1.6 \\
    12 & 1700 & 2300 & 1.0 & 1.8 \\
    14 & 2000 & 2600 & 1.2 & 2.0 \\
    16\tnote{\textasteriskcentered} & 2200 & 2800 & 1.4 & 2.2 \\
    18 & 2500 & 3100 & 1.5 & 2.4 \\
    20 & 2800 & 3400 & 1.7 & 2.6 \\
    \hline
    \end{tabular}
      \begin{tablenotes}
        \footnotesize
        \item[\textasteriskcentered] Actual number of days with major power outages in the Pacific Gas and Electric Company service territory in 2014 \cite{pge_reliability_2014}.
      \end{tablenotes}}
\end{threeparttable}

We calculate the marginal revenue earned serving power demand by taking the difference between each year and the annualized mobility-only scenario. Our results suggest that fleet operators can earn \$1,400-\$3,400 (or $\sim$1-3\%) additional revenue per vehicle per year serving power demand during outages, depending on fleet size and the number of major power outages.

These results are sensitive to numerous assumptions in our analysis, including but not limited to: outage cost, outage frequency/duration, outage size/scope, power demand, vehicle battery size, battery discharge rate, optimization window, and foresight into demand for power and passenger trips.

\section{Summary}
We demonstrate a method for simulating the energy and geospatial distribution of a fleet of autonomous PEVs in San Francisco dispatched to serve mobility and electricity demand during power outages. We use a PDE-based approach to model the aggregate state of energy of the fleet as vehicles charge, discharge, and travel throughout the system. We optimize vehicle dispatch over a 50 minute planning horizon, assuming perfect foresight into both mobility and power demand within that time frame. We consider two outage scenarios, including both Moderate and Extreme outages based on real outage data for San Francisco. Finally, we compute the revenue earned in each scenario with various fleet sizes, ranging from 7,500 to 40,000 vehicles. We find that serving power demand increases fleet revenue by \$1,400-\$3,400 per vehicle, or 30-40\%, in the Extreme outages scenario. Given that power outages are rare, these results translate to $\sim$1-3\% more revenue per year, depending on the number of major power outages in a year.

% use section* for acknowledgment
\section*{Acknowledgments}
The authors would like to thank Scott Moura and Caroline le Floch at UC Berkeley for advising this work. We also thank Michael Sohn and Joseph Eto at Lawrence Berkeley National Laboratory for granting us access to the outage dataset.

% trigger a \newpage just before the given reference
% number - used to balance the columns on the last page
% adjust value as needed - may need to be readjusted if
% the document is modified later
%\IEEEtriggeratref{8}
% The "triggered" command can be changed if desired:
%\IEEEtriggercmd{\enlargethispage{-5in}}

% references section

% can use a bibliography generated by BibTeX as a .bbl file
% BibTeX documentation can be easily obtained at:
% http://mirror.ctan.org/biblio/bibtex/contrib/doc/
% The IEEEtran BibTeX style support page is at:
% http://www.michaelshell.org/tex/ieeetran/bibtex/
\bibliographystyle{IEEEtran}
% argument is your BibTeX string definitions and bibliography database(s)
% \bibliography{refs}
%
% <OR> manually copy in the resultant .bbl file
% set second argument of \begin to the number of references
% (used to reserve space for the reference number labels box)

% that's all folks
\end{document}